\documentclass[PRL,twocolumn,superscriptaddress,showpacs,preprintnumbers,amsmath,amssymb,floats]{revtex4}

\usepackage{graphicx}

\begin{document}

\title{Inelastic X-Ray Scattering Study of Exciton Properties in an Organic Molecular Crystal}

\author{K.~Yang}
\affiliation{Department of  Physics, and Applied Surface Physics
State Key Laboratory, Fudan University, Shanghai 200433}
\author{L.~P.~Chen}
\affiliation{Key Laboratory of Organic Solids, Beijing National
Laboratory for Molecular Sciences (BNLMS), Institute of Chemistry,
Chinese Academy of Sciences, Beijing 100080}
\author{Y.~Q.~Cai}
\author{N.~Hiraoka}
\affiliation{National Synchrotron Radiation Research Center, Hsinchu
30076}
\author{S.~Li}
\author{J.~F.~Zhao}
\author{D.~W.~Shen}
\affiliation{Department of  Physics, and Applied Surface Physics
State Key Laboratory, Fudan University, Shanghai 200433}
\author{H.~F.~Song}
\author{H.~Tian}
\affiliation{Laboratory for Advanced Materials and Institute of Fine
Chemicals, East China University of Science and Technology, Shanghai
200237}

\author{L.~H.~Bai}
\author{Z.~H.~Chen}
\affiliation{Department of  Physics, and Applied Surface Physics
State Key Laboratory, Fudan University, Shanghai 200433}
\author{Z.~G.~Shuai} \email{dlfeng@fudan.edu.cn; zgshuai@iccas.ac.cn}\affiliation{Key Laboratory of Organic Solids, Beijing National
Laboratory for Molecular Sciences (BNLMS), Institute of Chemistry,
Chinese Academy of Sciences, Beijing 100080}
\author{D.~L.~Feng}
\email{dlfeng@fudan.edu.cn; zgshuai@iccas.ac.cn}
\affiliation{Department of Physics, and Applied Surface Physics
State Key Laboratory, Fudan University, Shanghai 200433}

\date{\today}

\begin{abstract}

Excitons in a complex organic molecular crystal were studied by
inelastic x-ray scattering (IXS) for the first time. The dynamic
dielectric response function is measured over a large momentum
transfer region, from which an exciton dispersion of 130meV is
observed. Semi-empirical quantum chemical calculations reproduce
well the momentum dependence of the measured dynamic dielectric
responses, and thus unambiguously indicate that the lowest Frenkel
exciton is confined within a fraction of the complex molecule. Our
results demonstrate that IXS is a powerful tool for studying
excitons in complex organic molecular systems. Besides the energy
position, the IXS spectra provide a stringent test on the validity
of the theoretically calculated exciton wavefunctions.

\end{abstract}

\maketitle

In the past several decades, optical applications of organic
materials have been extensively explored.
\cite{optical,pvc,tft,spiro,sensor,lumi,polymer} Organic light
emitting devices, photovoltaic cells, photochromic materials,
biosensors, and nonlinear optical devices among many others have
generated a lot of interests. Great effort has been invested on
designing new functional materials with special optical properties.

As the optical processes are largely dominated by excitonic
excitations in these materials, the detailed information of excitons
is crucial. Particularly, it would greatly benefit the material
design if one knows where the exciton is located in a molecule, what
its spatial extent is, and how the exciton energy and dispersion are
associated with the local structure.

These basic yet crucial questions on exciton cannot be addressed by
conventional techniques such as Raman scattering and absorption, as
they can only give information near zero momentum transfer. So far,
high energy electron-energy-loss spectroscopy (EELS) has been
applied successfully to thin film samples of conjugated oligomer or
polymer such as $\alpha$-$6T$, $6P$, carotenoid, and TPD.
\cite{fink99,fink00,fink04,quasiband}. EELS is limited to the low
momentum transfer $q$ due to strong multiple scattering at high $q$,
which would severely smear out the spectrum, and the diminishing
matrix element ($\sim q^{-4}$). As a result, EELS studies were
mostly focused on planar $\pi$ conjugated molecules where the
excitons are more extended in space, and information at low $q$ is
sufficient for understanding the exciton behaviors. On the other
hand, the majority of organic optical materials are composed of
small molecules with complex local structures. Even for polymers,
the optically functional portions are mostly small sectors attached
to the backbones for better mechanical performance. For small
molecules or small molecular clusters, the size of the excitons is
presumably very small, and high-$q$ information is crucial for
understanding their behaviors. The properties of excitons in these
systems remain largely unexplored experimentally. Theoretically,
although many quantum chemical methods have been employed to study
these systems, their feasibility lacks strong experimental support.

Inelastic X-ray scattering (IXS) has proven to be a powerful tool
for investigating the electronic excitations in inorganic systems.
\cite{schulkereview} For example, IXS data have helped our
understanding of the metal-insulator transition \cite{IssacsMI},
plasmon excitations \cite{shulkeplas}, and band gap \cite{bandgap}.
IXS is a clean and direct probe of the dynamic structure factor
$S(q,\omega)$, which is proportional to
$q^2\times{Im(\epsilon^{-1}})$, $\epsilon$ being the dynamic
dielectric function. In comparison with EELS, IXS is almost free
from multiple scattering effects, and it can generate reliable
information at high $q$. Therefore, IXS is particularly advantageous
for studying excitons that are more constrained in space.

In this paper, we report the first IXS measurement of the Py-SO
organic molecular crystal. The dynamic dielectric responses of such
complex small molecules were measured over a large momentum transfer
region. The dispersion of the lowest exciton was observed to be
about 130 meV. Quantum chemical calculations based on ZINDO
(Zerner's intermediate neglect of differential overlap)/SCI (single
configuration interaction)\cite{Ridley73} was performed on both
molecular aggregate and a single molecule. We found that the
ZINDO/SCI method captured very well the internal structure of the
excitons. As the momentum dependence of the dielectric function
gives strong constraints on the exciton wavefunctions, the good
agreement between the calculated IXS spectra and the experiments
indicates unambiguously that the lowest Frenkel exciton distributes
in a fraction of a single molecule. Our results provide
comprehensive information for understanding the properties of
excitons in such complex molecular systems.

\begin{figure}
\includegraphics[width=7cm]{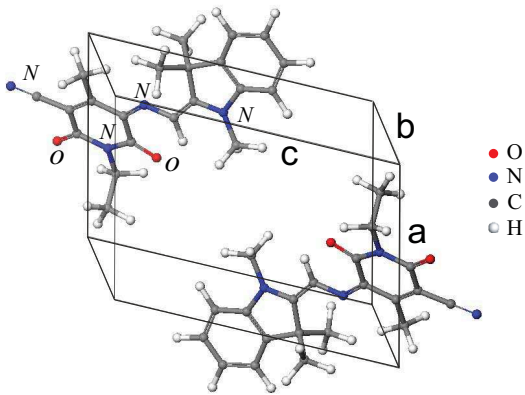}
\caption{A unit cell of
 Py-SO molecular crystal.}
 \label{structure}
\end{figure}

Spirooxazines (SO) are a type of well-known photochromic compounds.
Open-ring photomerocyanine form of spirooxazines (Py-SO,
C$_{21}$N$_4$O$_2$H$_{22}$) was synthesized recently\cite{pyso},
which exhibited alkali induced chromism and the thermochromism in
alkali medium. High quality single crystal Py-SO samples, typically
in the size of $0.8mm \times 2mm \times 150\mu m$, were prepared by
recrystallization. Fig.\,\ref{structure} shows one unit cell, which
contains two molecules. The bottom right Py-SO molecule, from left
to right, consists of one phenyl ring, one aromatic ring, and a C-N
bond bridge to the quinoid on the right. At room temperature, the
triclinic lattice parameters are $a = 8.356 $\r{A}, $b = 9.510 $\AA,
$c = 12.096 $\AA, $\alpha =89.93^o$, $\beta =75.58^o$, $\gamma
=81.92^o$.

IXS measurements were carried out in transmission mode at the Taiwan
Beamline BL12XU at SPring-8.\cite{CaiExp}  A Si(444) spherical
analyzer with 2m radius of curvature was used. A Si(400)
high-resolution monochromator was used to scan incident photon
energy around 7908.75eV. The total energy resolution was estimated
to be about 170meV based on the FWHM of the quasi-elastic lines of
the sample. Momentum resolution was about 0.14\AA$^{-1}$. Data were
taken at room temperature. Great care was taken to avoid beam damage
to sample during the experiments by changing the probing spot every
scan, since the beam spot was only about $ 120\mu m \times 80\mu m$.
The time for one scan was less than 4 hours, and the damage effects
were negligible. Data count rate was about 10Hz. Absorption is only
about $6\%$, therefore it is neglected in the following analysis.

\begin{figure*}
\includegraphics[width=17.5cm]{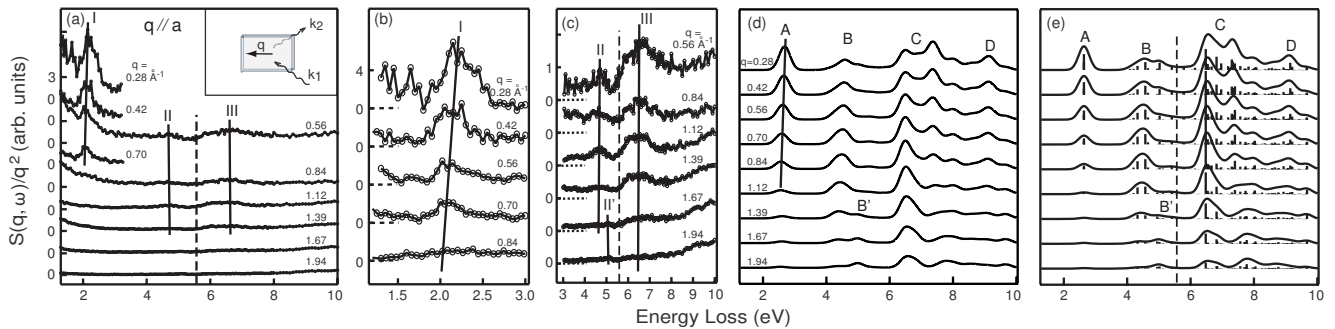}
\caption{(a) IXS data of Py-SO with momentum transfer $q$ along the
$a$ axis. The spectral intensities are divided by $q^2$. (b, c)
Enlargements of spectra shown in (a) with elastic line removed.
ZINDO/SCI simulated $S(q,\omega)/q^2$ for Py-SO for $q$ along the
$a$ axis based on (d) the aggregate and (e) a single molecule. The
dashed lines in panels a,c, and e indicate the calculated HOMO-LUMO
gap energy.} \label{spectra}
\end{figure*}

$Im(\epsilon^{-1})$ data of the Py-SO molecular crystal measured by
IXS are shown in Fig.\,\ref{spectra}a with momentum transfer
\emph{q} along the $a$ axis. The spectral features are determined by
the internal structure of the optically excited singlet excitons and
interband transitions. There are three distinct features at about
2.2eV, 4.6eV and 6.6eV in the measured energy window, labeled as I,
II, and III respectively. After removing the quasi-elastic Rayleigh
background, the resulting spectra are shown in
Fig.\,\ref{spectra}b-c. A weak feature II' and rising spectral
weight beyond 8 eV can also be observed in Fig.\,\ref{spectra}c.

Exciton, or excited state in general has been a major challenge for
computational physics or chemistry because of the difficulty in
treating electron correlation for excited states. Hutchison and
coworkers have made a systematic investigation for the excited
states of 60 organic conjugated molecules with six commonly applied
excited-state computational methods. When compared with experiments,
ZINDO/SCI combined with Austin-Model-1-optimized geometry was found
to be the best choice in predicting the low-lying excited states,
even outperforming the most commonly applied first-principles
time-dependent density functional theory.\cite{Hutchison02}
Therefore it is adopted in our calculations.

The IXS spectra calculated along the $a$ axis are shown in
Fig.\,\ref{spectra}d for a molecular aggregate of six-unit-cell
stack with shortest intermolecular separations, \cite{aggregate} and
a width of 0.17eV was included in order to compare with the
experiment. One can also clearly identify several main spectral
features A, B and B$^\prime$, C and D. There are almost one to one
correspondence between the experimental features I-III and
theoretical features A-C, respectively. The energy centroid
positions of features B and C match those of features II and III
almost perfectly. The weak feature B$^\prime$ at about 5eV becomes
pronounced from 1.12\AA$^{-1}$ and above. The corresponding feature
II$^\prime$ becomes visible at high momentum transfers when feature
II is weak. Both feature C and D involve many excitations.
Correspondingly, the spectra in the experiment exhibit very broad
feature III followed with rising spectral weight. These qualitative
and, to a great extent, quantitative agreements between the
theoretical spectra and the experimental results indicate that
ZINDO/SCI calculations characterize the Py-SO system quite well.

The experimental feature I disperses from 2.2eV at
$q=0.28$\AA$^{-1}$ to 2.07eV at $q=0.7$\AA$^{-1}$. This  is well
reproduced in the theoretical spectra, where the peak position of A
disperses by 0.12eV in the corresponding momentum range, except that
the theoretical position is about 0.48eV higher. The small exciton
dispersion reflects the weak intermolecular coupling.
\cite{quasiband} In the aggregate calculation, the strongest
intermolecular coupling between various orbitals in neighboring
molecules is estimated to be 55meV. Therefore, the excitons are
still Frenkel excitons that are confined mostly in a single molecule
in this case. As a result, single molecule excitations could be
computed to study the local distribution of the electrons and holes.
In fact, $S(q,\omega)$ from single molecule calculations has been
shown to agree very well with the EELS measurements for conjugated
oligomers and polymers.\cite{Shuai98} The IXS spectra calculated for
molecular excitations of single Py-SO molecule along the $a$ axis
are shown in Fig.\,\ref{spectra}e , where energy levels of the
excited states are indicated by straight lines. As expected, the
spectra are very similar to the aggregate calculations. The energy
gap between LUMO and HOMO orbitals, is calculated to be 5.62 eV in
ZINDO/SCI, denoted by the dashed line in Fig.\,\ref{spectra}.
Moreover, our calculation shows that the lowest energy feature A
corresponds to a discrete exciton, unlike in long chain systems,
\cite{mukamelprl,mukameljcpa} where the peaks in the spectra might
be consisted of many excitons. On the other hand, feature B contains
mostly two major excitons and feature C and D are made up of tens of
excitations above the gap. In experiment, as individual excitations
could not be distinguished in features II and III, their dispersion
could not be resolved even if there is any.

\begin{figure*}
\includegraphics[width=12.5cm]{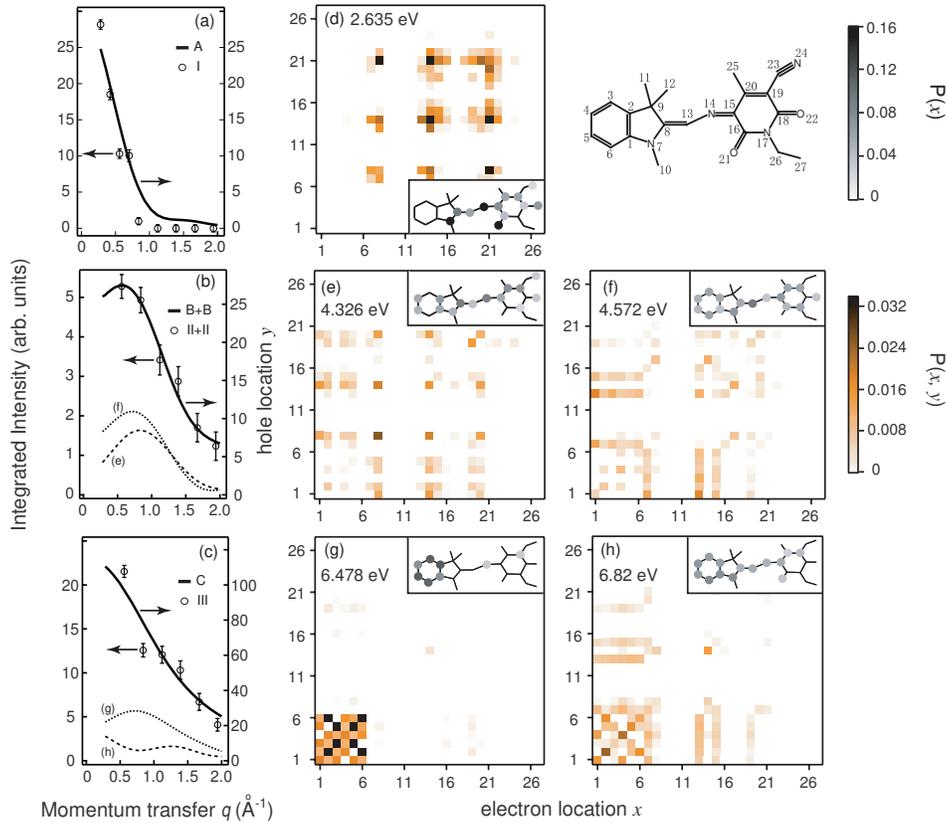}
\caption{Integrated intensity as a function of momentum transfer for
experimental features and single molecule calculations: (a) A and I
integrated over [1.5eV, 3eV], (b) B+B' and II+II' integrated over
[4eV, 5.4eV], and (c) C and III integrated over [5.4eV, 8eV]. Broken
and dashed lines in (b) and (c) show the q dependence of two main
excitations of features B and C respectively. The false color plots
for the calculated possibility $P(x,y)$ of finding an electron at
site $x$ and a hole at site $y$ for (d) the lowest energy exciton;
(e,f) two main excitations of feature B; (g,h) two main excitations
of feature C. The atom sites are numbered in the top right corner.
The atoms plotted in gray scale on the Py-SO molecule structure in
the inset represent the probability $P(x)\equiv\sum_{y}P(x,y)$ of
finding the electron or hole at site $x$ for Py-SO molecule.}
\label{orbital}
\end{figure*}

As the dynamic structure factor is \emph{directly} determined by the
ground state and excited state wavefunctions, it puts stronger
constraints on theoretical exciton distributions than just exciton
energy position information derived from conventional optical
measurements. The momentum dependence of the integrated weight is
compared with the single molecule calculation in
Fig.\,\ref{orbital}a-c for the three main features respectively.
After rescaling, the theory agrees with the experiments very well.
However, as shown in Fig.2, the theoretical high energy features are
more pronounced than the experiment, producing higher integrated
weights in Fig. 3b-c compared to feature A. We attribute this
overestimation to the more delocalized nature of the high energy
excitons, which are less bound than the lowest exciton. Therefore,
our single molecule calculation would overestimate its on-site
occupation, which in turn causes higher matrix element of IXS.
Indeed, the current aggregate calculation does give lower integrated
weight, but more detailed calculation on larger aggregate is needed
to fully address this issue. As the large intermolecular distance
corresponds to a momentum smaller than the sampled range, the
calculated local distribution of the high energy excitons in a
single molecule could still be compared with the experimental data.
For further confirmation, IXS spectra were sampled at two momentum
transfers perpendicular to the $a$ direction in the $ac$ plane,
which also shows good agreement with the theory.(Fig.\,4) The fact
that the theory based on a single molecule matches the experiment so
well indicates that the exciton distribution within the molecule is
well captured, as a result of the weak coupling between the
molecules. Fig.\,\ref{orbital}d-h are exciton wavefunction presented
in a way that the false color scale indicates the possibility
$P(x,y)$ for finding an electron at atom site $x$ and a hole at atom
site $y$. The gray scale of the solid circles in the inset shows the
possibility $P(x)\equiv\sum_{y}P(x,y)$ of finding the electron or
hole at site $x$ on Py-SO molecule. These plots give clear
information on how the electron and hole of a certain exciton are
distributed in the molecule. For feature A, exciton is mostly
situated in the middle region. For feature B, on the other hand,
both of the two main excitons are extended over the entire molecule,
which are much larger than the lowest energy exciton.(Fig.3e-f)  For
comparison, the two main inter-gap excitations of feature III are
shown in Fig.3g-h, one being confined in the phenyl ring, the other
being extended.

\begin{figure}[b!]
\includegraphics[width=8.5cm]{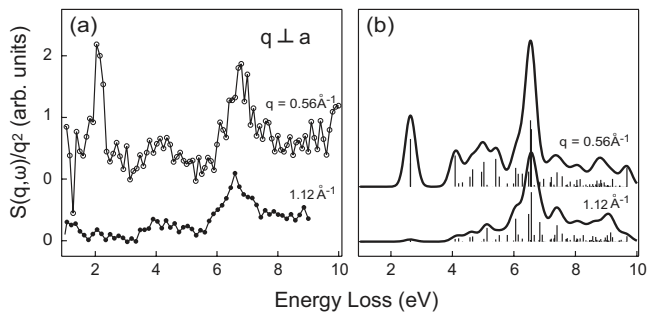}
\caption{(a)Experimental data with background removed. Momentum
transfer $q$ is perpendicular to $a$ axis. (b) ZINDO/SCI simulated
$S(q,\omega)/q^2$ based on a single molecule with $q$ perpendicular
to $a$ axis in the $ac$ plane} \label{f4}
\end{figure}

The measured intensity distribution of $S(q,\omega)$ corresponds to
the distribution of the exciton center-of-mass in the system. For
simple linear molecular systems such as oligomer and polymer,
$S(q,\omega)$ along the chain direction relates directly with the
exciton distribution over the molecule. The situation is more
complicated for a molecular system with complex internal structures
such as Py-SO. The structure of $S(q,\omega)$ measured along a
certain $\hat{q}$ direction reflects distribution of the exciton
center-of-mass in this direction \emph{summed} across the entire
molecule. Because the momentum transfer does not correspond to the
relative electron-hole motion, quantum chemical calculations is
needed to retrieve the full quantitative details of the exciton
wavefunctions.\cite{mukamelprl} Nevertheless, there are still some
qualitative correspondence. For example, the significant occupation
of exciton A on atomic sites \#15,16,19,20,21,24 (Fig.3d) generally
makes its center-of-mass more delocalized than the others along
$\hat{a}$-direction (Fig.1), and thus the momentum distribution of
$S(q,\omega)$ is narrower in Fig.3a.

Inelastic x-ray scattering is a weak probe. From the experimental
point of view, this is the first time that IXS is demonstrated to be
feasible for organic molecular crystal in a third generation
synchrotron. The experimental results are very clean and therefore
can be directly compared with theory. The dispersion of the exciton
feature gives a good measure of the strength of the intermolecular
coupling. Combined with suitable quantum chemical calculations,
reliable and comprehensive properties of excitons can be obtained,
which are crucial for understanding their optical properties, and
for designing materials of desired optical properties based on
exciton transfer or dissociation properties. For example, if
excitons in this useful energy range (2.2eV here) are localized
within a particular complex structure, molecular clusters with a
similar structure might be attached to polymers without affecting
their local exciton (optical) behavior.

{\it Acknowledgements:} DLF would like to thank Profs. G. A
Sawatzky, X. Sun, C. Q. Wu, and H. Chen for very stimulating
discussions. This work was supported by the NSFC, and the 973
Project of MOST of China(Grant No: 2002CB613406), and by the
Shanghai Science and Technology Committee. Experiments at SPring-8
were partially supported by the NSC of Taiwan (Grant No.:
NSC94-2112-M-213-012). The computation has been carried out in the
CNIC Supercomputing Center of the CAS.

\end{document}